\documentclass[a4paper,11pt]{article}
\usepackage{pos}

\title{Background measurements and detector response studies for ISMRAN experiment}

\author*[a,c]{R.Dey}
\author[a]{P.~K.~Netrakanti}
\author[a]{D.~K.~Mishra}
\author[a]{S.~P.~Behera}
\author[a]{R.~Sehgal}
\author[a,c]{V.~Jha}
\author[b,c]{L.~M.~Pant}

\affiliation[a]{Nuclear Physics Division, Bhabha Atomic Research Centre,\\
  Trombay, Mumbai - 400085, India.}

\affiliation[b]{Technical Physics Division, Bhabha Atomic Research Centre,\\
Trombay, Mumbai - 400085, India.}

\affiliation[c]{Homi Bhabha National Institute,\\
Anushakti Nagar, Mumbai - 400094, India.}

\emailAdd{rnd48388@gmail.com}

\abstract{We report the measurement of the non-reactor environmental backgrounds and the detector response with the Indian Scintillator Matrix for Reactor Anti-Neutrinos (ISMRAN), which is $\sim$1 ton detector setup by volume, consisting of 10$\times$9 (10 rows and 9 columns) Plastic Scintillator Bars (PSBs) array at BARC, Mumbai, India. ISMRAN is an above-ground anti-neutrino (${\overline{\ensuremath{\nu}}}_{e}$) experiment at very short baseline located at Dhruva research reactor facility. It is enclosed by a shielding made of 10 cm thick lead and 10 cm thick borated polyethylene to minimize the backgrounds and is mounted on a movable base structure, situated at $\sim$ 13 m away from the reactor core. These measurements are useful in the context of the ISMRAN detector setup that will be used to detect the reactor ${\overline{\ensuremath{\nu}}}_{e}$ and measure its energy spectrum through the inverse beta decay (IBD) process. In this paper, we present the energy resolution model and energy non-linearity model of PSB and the cosmogenic muon-induced background, based on the sum of their energy depositions and number of hit bars. Reconstructed sum energy spectrum and number of hit bars distribution for $\mathrm{{}^{22}Na}$ radioactive source has been compared with Geant4 based Monte Carlo simulations. These experimentally measured results will be useful for discriminating the correlated and uncorrelated background events from the true IBD events in reactor ON and OFF conditions inside the reactor hall.}

\FullConference{%
  41st International Conference on High Energy physics - ICHEP2022\\
  6-13 July, 2022\\
  Bologna, Italy
}

\begin{document}
\maketitle

\vspace*{-15mm}

\section{Introduction:}
\begin{figure}
\centering
\includegraphics[scale=.24]{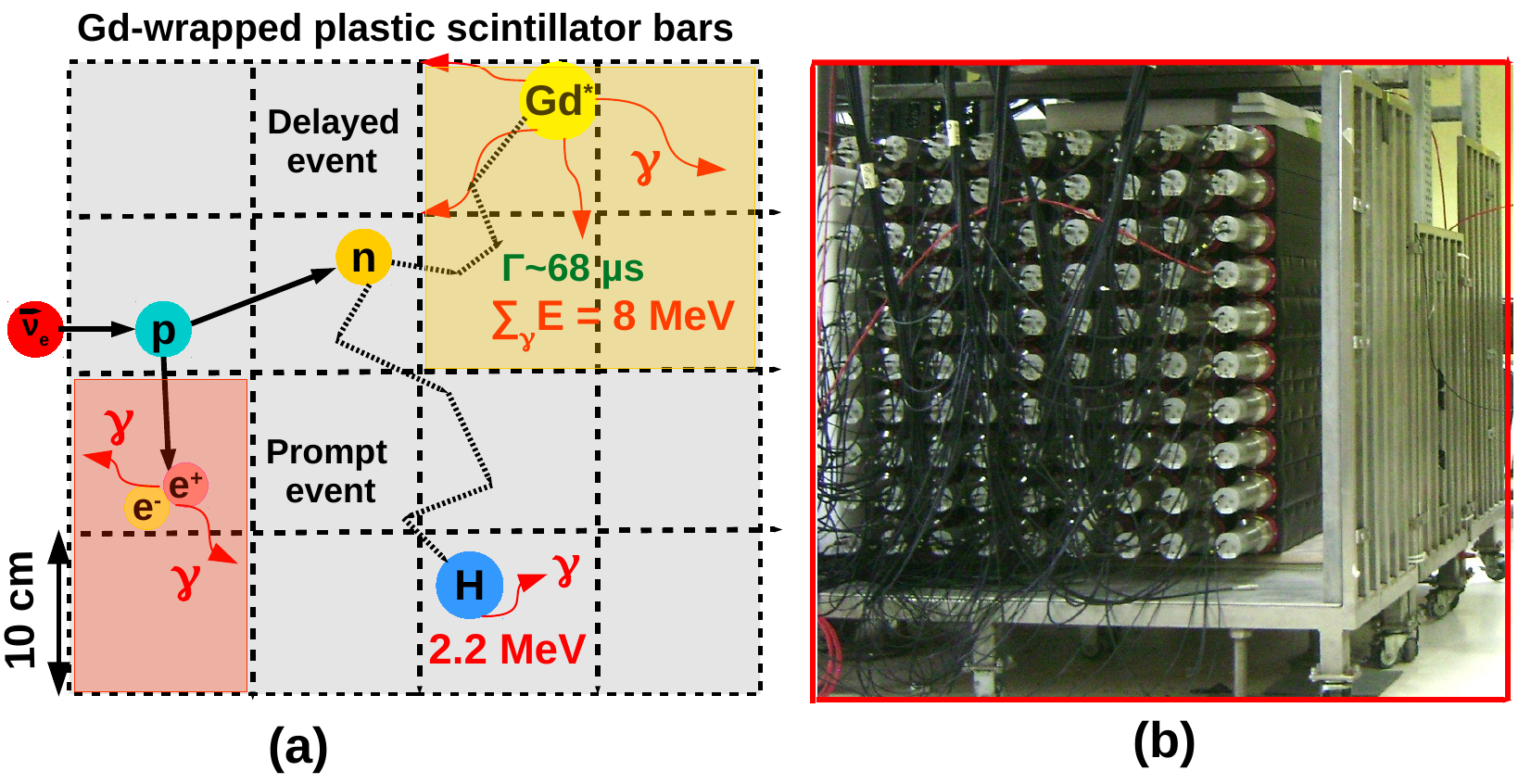}
\caption{(a) Schematic representation of IBD event generating prompt and delayed event signatures in a prototype mini-ISMRAN array. (b) Full scale ISMRAN detector setup consisting of 90 PSBs in DIL. }
\label{fig1}  
\end{figure}

Indian Scintillator Matrix for Reactor Anti-Neutrinos (ISMRAN) detector setup was designed to measure the yield and energy spectrum of ${\overline{\ensuremath{\nu}}}_{e}$, via the inverse beta decay (IBD) ($\mathrm{ \overline \nuup_{e} + p \rightarrow e^{+} + n}$) process which is shown in Fig~\ref{fig1} (a), for monitoring the reactor thermal power, fuel evolution in a non-intrusive way and also searching for the existence of light sterile neutrino. The detector setup at Detector Integration Laboratory (DIL) in BARC, consists of 90 PSBs, arranged in an array of 10$\times$9 which is shown in Fig~\ref{fig1} (b) in non-reactor environment. Each PSB is wrapped with Gadolinium Oxide ($\mathrm{Gd_{2}O_{3}}$) coated aluminized mylar foils and is 100 cm long with a cross-section of 10$\times$10$~\mathrm{cm^{2}}$ ~\cite{roni}. Three-inch diameter PMTs are coupled at both ends of each PSB. The data acquisition system, a CAEN VME-based 16 channels and 14 bits waveform digitizer (V1730) with high sampling frequency of 500 MS/s has been used for the pulse processing and event triggering from each PSB independently. The anode signals from PMTs at both ends of a PSB are required to have a time coincidence of 20 ns to be recorded as a triggered event. 





\vspace*{-2mm}
\vspace*{-2mm}
\section{Detector response and backgound measurements in non-reactor environment:}
A good understanding of the response of the PSB is an essential pre-requisite for the analysis presented in this paper. In MC simulations based on GEANT4, the standard electromagnetic and radioactive decays physics processes are used for the response of $\gamma$-rays, electrons and positrons from radioactive sources. 




\begin{figure}
\centering
\includegraphics[scale=.65]{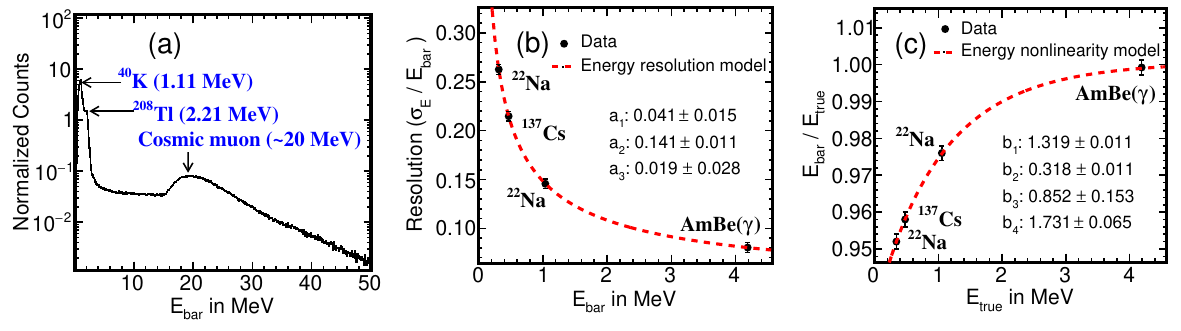}
\caption{(a) Calibrated $\mathrm{E_{bar}}$ distribution in a single PSB for the non-reactor background, (b) energy resolution as a function of reconstructed $\gamma$-ray energy and (c) non-linear response of scintillating energy in PSB obtained from different radioactive sources.}

\label{fig2}  
\end{figure}

\begin{equation}\label{eq:Resolution}
  \hspace{-0.2in}
\mathrm{ Resolution \left (\frac{\sigma_{E}}{E_{bar}} \right ) = \sqrt{a_{1}^{2} + \frac{a_{2}^{2}}{E_{bar}} + \frac{a_{3}^{2}}{E_{bar}^{2}}}},
\end{equation}

\begin{figure}
\centering
\includegraphics[scale=.65]{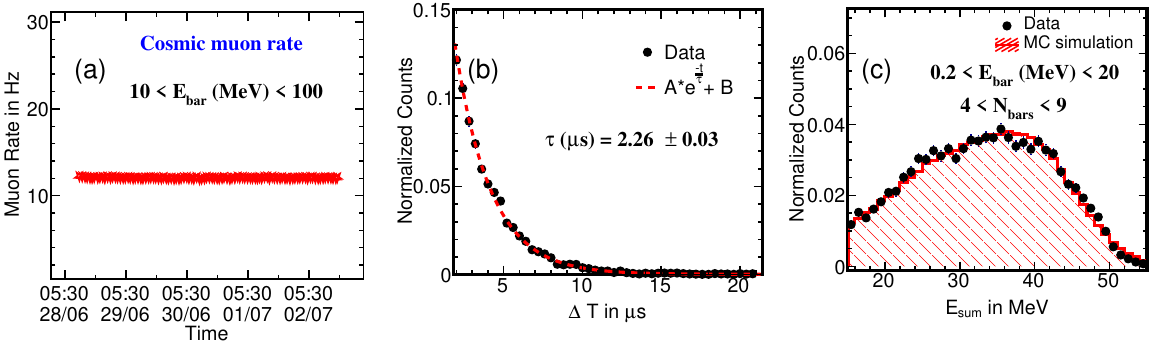}
\caption{(a) Cosmic muon rate measured at sea lavel as a function of time, (b) distribution of the measured cosmic muon decay time (solid) fit with an exponential function (dashed) and (c) comparison between data and MC for the reconstructed sum energy ($\mathrm{E_{sum}}$) in delayed events (Michel electron).}
\label{fig3}  
\end{figure}

Figure~\ref{fig2} (a) shows the energy deposition ($\mathrm{E_{bar}}$) in a single PSB due to the non-reactor backgrounds. Natural radioactive background mainly dominated by the Compton edges of $\mathrm{{}^{40}K}$ ($\mathrm{E_{bar}}$ = 1.11 MeV) and $\mathrm{{}^{208}Tl}$ ($\mathrm{E_{bar}}$ = 2.21 MeV) are below 3 MeV and the feature at $\sim$ 20 MeV is due to the minimum ionization energy deposited by cosmic muons in the 10 cm thick PSB. Depending on their $\mathrm{E_{bar}}$ in PSB, non-reactor backgrounds have been separated into two regions, (A) signal region $\mathrm{(0.2 < E_{bar} (MeV) < 10.0)}$ and (B) cosmic muon background region $\mathrm{(10.0 < E_{bar} (MeV) < 50.0)}$. Figure 2(b) displays the energy dependent resolution model with reconstructed $\gamma$-ray energy for $\mathrm{{}^{137}Cs}$, $\mathrm{{}^{22}Na}$ and $\gamma$-rays from Am/Be source in a single PSB. These results indicate that energy resolution of ISMRAN is well described by the standard resolution formula, which is shown in Eqs.~\ref{eq:Resolution} and that the photo-statistics term is the primary component of the energy resolution for ISMRAN, which is $\sim$14$\%$/$\sqrt{\mathrm{E_{bar}}}$ ~\cite{roni}. Figure~\ref{fig2} (c) shows the nonlinear response of scintillating energy in a single PSB for different radioactive $\gamma$-rays sources, that is well described by a fitted parametrization and consistent with the MC predictions. The nonlinear response at lower energies is mainly due to the quenching effect in the scintillator volume. The following empirical formula is used for the description of the energy non-linearity of PSB:

\begin{equation}\label{eq:Nonlinearity}
\mathrm {\frac{E_{bar}}{E_{true}} = b_{1} - \frac{b_{2}}{[1 - e^{-(b_{3}E_{true} + b_{4})}]}}
\end{equation}





Figure ~\ref{fig3} (a) shows the measured rate of cosmic muon over the period of time. The obtained average cosmic muon flux is $\sim 0.9$ $\mathrm{cm^{-2}}$$\mathrm{min^{-1}}$ for the $\mathrm{E_{bar}}$ range $\mathrm{10.0 < E_{bar} (MeV) < 50.0}$ at sea level, which is consistent with the measured results of other experiment within the statistical errors. In the non-reactor environment, correlated background is mainly dominated by stopping muons (SM). The selected SM arise from the muons entering through the top layers of PSBs in ISMRAN array, stopping inside the array and eventually decaying into $e^{-}$ or $e^{+}$ (Michel electrons) via the following processes:

\begin{equation}\label{eq:Michel electron1}
\mathrm{ \mu^{-} \rightarrow e^{-} + \nuup_{\mu} + \overline \nuup_{e} }
\end{equation}
\begin{equation}\label{eq:Michel electron2}
\mathrm{ \mu^{+} \rightarrow e^{+} + \nuup_{e} + \overline \nuup_{\mu} }
\end{equation}

Figure ~\ref{fig3} (b) shows the $\Delta T$ distribution of prompt-delayed pairs for SM events inside the ISMRAN array. The $\Delta T$ distribution is fitted with a combined function consisting of an exponential term and a constant term representing the accidental background. For SM events, the fit results in the characteristic time ($\tau$) of 2.26 $\pm$ 0.03 $\mu$s which is in good agreement with the theoretical estimation of mean lifetime of cosmic muon $\sim$ 2.2 $\mu$s at sea level. Figure ~\ref{fig3} (c) describes the reconstructed sum energy ($\mathrm{E_{sum}}$) distribution for delayed Michel electron candidate events. $\mathrm{E_{sum}}$ has been reconstructed for the delayed events following SM, $\mathrm{E_{bar}}$ is required to be in the range of 0.2 to 20.0 MeV, the number of hit bars ($\mathrm{N_{bars}}$) should be greater than 4 and less than 9 and $\mathrm{E_{sum}}$ should be 15 MeV to 60 MeV within the characteristic time window of 1 $\mu$s to 15 $\mu$s. Michel electrons from cosmic muons decays provide a cross-check of the energy calibration for our ISMRAN detectors.

 \vspace*{-4mm}
\section{Reconstruction of $\mathrm{N_{bars}}$ and $\mathrm{E_{sum}}$ for $\mathrm{{}^{22}Na}$ source in the ISMRAN array:}

\begin{figure}
\centering
\includegraphics[scale=.45]{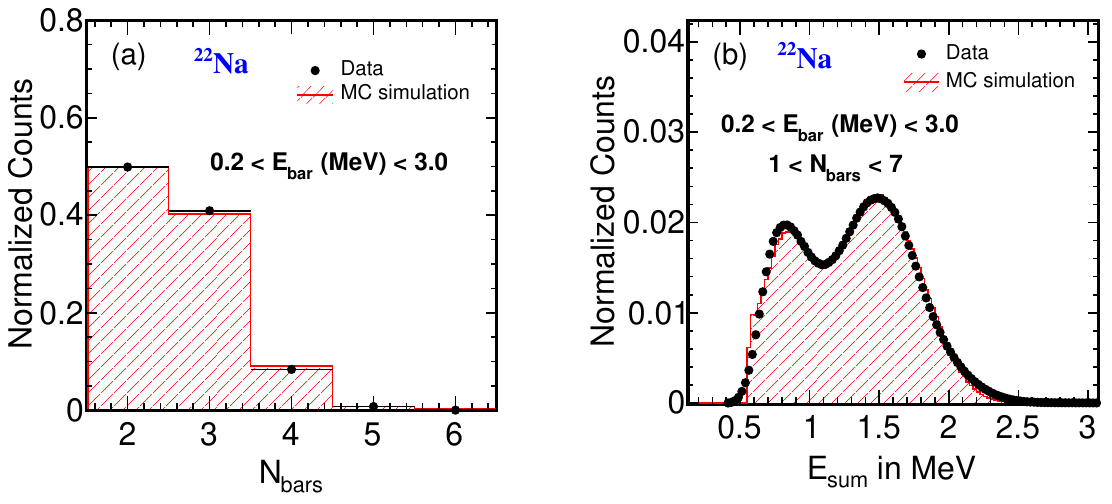}
\caption{Comparison of measured and simulated (a) $\mathrm{N_{bars}}$ and (b) $\mathrm{E_{sum}}$ distributions for $\mathrm{{}^{22}Na}$ source placed at the center of ISMRAN array.}
\label{fig4}  
\end{figure}

Figure ~\ref{fig4} (a) and (b) show the comparison of reconstructed $\mathrm{N_{bars}}$ and $\mathrm{E_{sum}}$ distributions in data and MC simulation for $\mathrm{{}^{22}Na}$ source placed at the center of the ISMRAN array. Only those PSBs are selected for the reconstruction of $\mathrm{N_{bars}}$ and $\mathrm{E_{sum}}$ where the individual energy deposition in each PSB is between 0.2 MeV to 3.0 MeV, so as to minimize the cosmic muon background. There is reasonably good agreement between data and MC results for the $\mathrm{N_{bars}}$ and $\mathrm{E_{sum}}$ distributions. As it can be seen from Fig ~\ref{fig4}(b), a peak at $\sim$1.6 MeV corresponds to the coincidence of $\gamma$-ray and positron events from $\mathrm{{}^{22}Na}$ source and a feature at $\sim$0.9 MeV corresponds to the $\gamma$-ray of energy 1.274 MeV originating from $\mathrm{{}^{22}Na}$ source in $\mathrm{E_{sum}}$ distribution, which has been reconstructed in the range of $\mathrm{1 < N_{bars} < 7}$.

\vspace*{-4mm}
\section{Summary:}
An energy resolution of $\sim$14$\%$ at 1 MeV has been achieved for the PSBs. We have also measured the integrated cosmic muons flux at sea level as a function of time and cosmic muon induced backgrounds based on their $\mathrm{E_{bar}}$ and $\mathrm{N_{bars}}$ using the ISMRAN detector setup. $\mathrm{E_{sum}}$ and $\mathrm{N_{bars}}$ variables have been validated by comparing with GEANT4 based MC simulations for radioactive $\gamma$-ray + positron source such as $\mathrm{{}^{22}Na}$ placed at the center of the ISMRAN array. The full scale ISMRAN experiment has been installed and commissioned in the Dhruva reactor hall on a movable base structure. The physics data campaign started at the end of 2021.


\end{document}